\documentclass[journal]{IEEEtran}
\usepackage{ifpdf}
\usepackage{cite}
\usepackage{amsmath, amssymb}
\usepackage{algorithm}
\usepackage{algorithmic}
\usepackage{multirow}
\usepackage{array}
\usepackage[pdftex]{graphicx}
\usepackage{subcaption}

\usepackage{dblfloatfix}
\usepackage{url}
\usepackage[dvipsnames]{xcolor}
\usepackage{tabularx}
\usepackage{diagbox}

\begin{document}
\title{Predictive Reliability Assessment of Distribution Grids with Residential Distributed Energy Resources}

\author{Arun~Kumar~Karngala,~\IEEEmembership{Student~Member,~IEEE,}
        Chanan~Singh,~\IEEEmembership{Life~Fellow,~IEEE,}
        and~Le~Xie,~\IEEEmembership{Fellow,~IEEE}% <-this % stops a space
\thanks{A. Karngala, C. Singh, and L. Xie are with the Department of Electrical and Computer Engineering, Texas A\&M University, College Station. L. Xie is with Harvard John A. Paulson School of Engineering and Applied Sciences.}% <-this % stops a space
}
\maketitle

\begin{abstract}
The growing adoption of behind-the-meter distributed energy resources (DERs) such as rooftop photovoltaic (PV) systems and energy storage (ES) is transforming end users from passive consumers to active participants in distribution systems, creating new challenges for reliability evaluation. This paper proposes a bottom-up probabilistic framework that integrates residential DER adoption into predictive reliability assessment. The framework models PV and ES adoption using joint probability distributions at the customer level, incorporates component reliability, and employs an adaptive Monte Carlo simulation to estimate not only mean values of reliability indices but also their distributional variability across customers. The approach is demonstrated on the RBTS Bus 4 system under sixteen joint adoption scenarios. Results show that high PV–storage adoption can reduce system-average SAIFI and SAIDI by more than 90\% compared to the baseline, while shallow adoption may worsen interruption frequency for some users. These findings highlight the importance of capturing adoption heterogeneity and distributional outcomes in reliability assessment, providing utilities and regulators with a robust tool for planning in DER-rich distribution grids.
\end{abstract}


% \fi
\IEEEpeerreviewmaketitle

\section{Introduction}
\IEEEPARstart{D}{istribution} systems have transitioned from unidirectional, static, radial systems~\cite{ch7textalan96} to bi-directional, dynamic, and meshed systems, evolving into complex interconnected networks. The importance of distribution system reliability for utilities and distribution system operators (DSOs) cannot be overstated, encompassing areas from community welfare to public health and economic vitality~\cite{Lacey2010TheDS}. Utilities increasingly apply asset management techniques to their distribution systems to identify factors influencing reliability  and to make informed decisions about maintenance, upgrades, and investments~\cite{adefarati2019reliability}.  

Customer satisfaction, an integral measure for utilities, is directly influenced by system reliability. Utilities ensure a consistent power supply by assessing reliability and fostering an environment conducive to economic development and heightened customer satisfaction. Reliability assessments also help utilities optimize infrastructure investments, minimizing the costs of interruptions to both utilities and customers, and thereby contributing to a more cost-effective and efficient distribution system~\cite{elkadeem2019optimal}.  

Two forms of reliability assessment are commonly applied. Historical assessment evaluates realized system performance using past and current data, while predictive assessment quantifies potential reliability outcomes under projected changes, scenarios, or future contingencies. Predictive assessment is particularly critical for planning, as it enables system designs that can adapt to emerging challenges. This paper is motivated by the need for predictive reliability methods that explicitly integrate end-user distributed energy resources (DERs), which are reshaping the nature of distribution system performance.  

\subsection{Related Works}

\textbf{Analytical approaches.} 
Early research on distribution system reliability with DERs emphasized analytical formulations. Examples include generalized reliability evaluation with distributed generation and microgrids~\cite{ContiMG}, encoded Markov set algorithms for hybrid renewable--conventional DG systems~\cite{Heydt13}, and methods explicitly considering protection strategy, islanding, and restoration in the presence of DG~\cite{chanananalytical}. These methods provide closed-form insights and computational efficiency but are typically limited by state-space growth and difficulty in capturing the stochastic dynamics of intermittent DERs at scale. More recent analytical studies~\cite{Zou2017, Wang2018, Wu2021} have extended recursive and state enumeration formulations to active distribution networks, microgrids, and DER uncertainty. While these models improve tractability for certain contexts, they continue to focus on mean load-point or system indices and cannot capture heterogeneity introduced by diverse end-user adoption.  

\textbf{Simulation approaches.} 
Monte Carlo simulation frameworks provide a more flexible alternative, enabling sequential and non-sequential representations of DER operation. Sequential simulation has been used to evaluate energy storage adequacy and microgrid islanding~\cite{xu2012adequacy, celli2013reliability}, while non-sequential methods incorporate DER and EV adoption under diverse outage scenarios~\cite{rocha2016reliability, WangPHEV, zhang2019reliability}. 
More recent works~\cite{Li2016customer, Anand2020, Song2022} have applied MCS to customer reliability, PHEV integration, and demand response. However, these studies largely report average indices such as SAIFI and SAIDI at the feeder or system level, and seldom explore the distributional variability across customers or include component reliability of PV and storage devices. 

In summary, analytical approaches offer compact formulations but suffer from state-space limitations and difficulty in modeling stochastic DER adoption, 
while simulation approaches provide flexibility but are computationally demanding and typically report only average reliability indices. Neither category fully captures the distributional impacts of heterogeneous end-user DER adoption or integrates component-level PV and storage reliability, 
which motivates the need for a new framework developed in this work.

\subsection{Motivation and Scope of this Work}

\textbf{Motivating example: [DSO vs. customer reliability]}  
Behind-the-meter (BTM) DERs allow end users to act as both sinks and sources of power depending on operating conditions. This fundamentally changes how reliability should be measured. Consider a simple distribution system with two load points, each serving five residential customers, as summarized in Table~\ref{tab:exampleLP}.  

\begin{table}[!t]
    \centering
    \caption{Load point data for example system}
    \label{tab:exampleLP}
    \begin{tabular}{cccc}
    \hline
         \textbf{Load point} & \textbf{Customers} & \textbf{$\lambda$ (f/yr)} & \textbf{U (h/yr)}  \\
         \hline
         LP1 & [1,2,3,4,5]   & 3 & 5  \\
         LP2 & [6,7,8,9,10] & 2 & 10 \\
         \hline
    \end{tabular}
\end{table}

Using the traditional approach, reliability indices are evaluated with the basic load point data as
\begin{equation}
    \text{SAIFI}_{P} = \frac{\sum \lambda_i C_i}{\sum C_i}, 
    \quad 
    \text{SAIDI}_{P} = \frac{\sum U_i C_i}{\sum C_i},
\end{equation}
where the subscript ``P'' denotes \emph{perceived} indices as reported by the DSO/utility.  
The actual customer experience can be represented by
\begin{gather}
    \text{SAIFI}_{E} = \frac{ \sum_C \text{Interruptions experienced}}{\sum C}, \\
    \text{SAIDI}_{E} = \frac{ \sum_C \text{Interruption duration experienced}}{\sum C},
\end{gather}
where the subscript ``E'' denotes \emph{experienced} indices at the end-user level.  

\begin{figure}[!t]
    \centering
    \includegraphics[width=\linewidth]{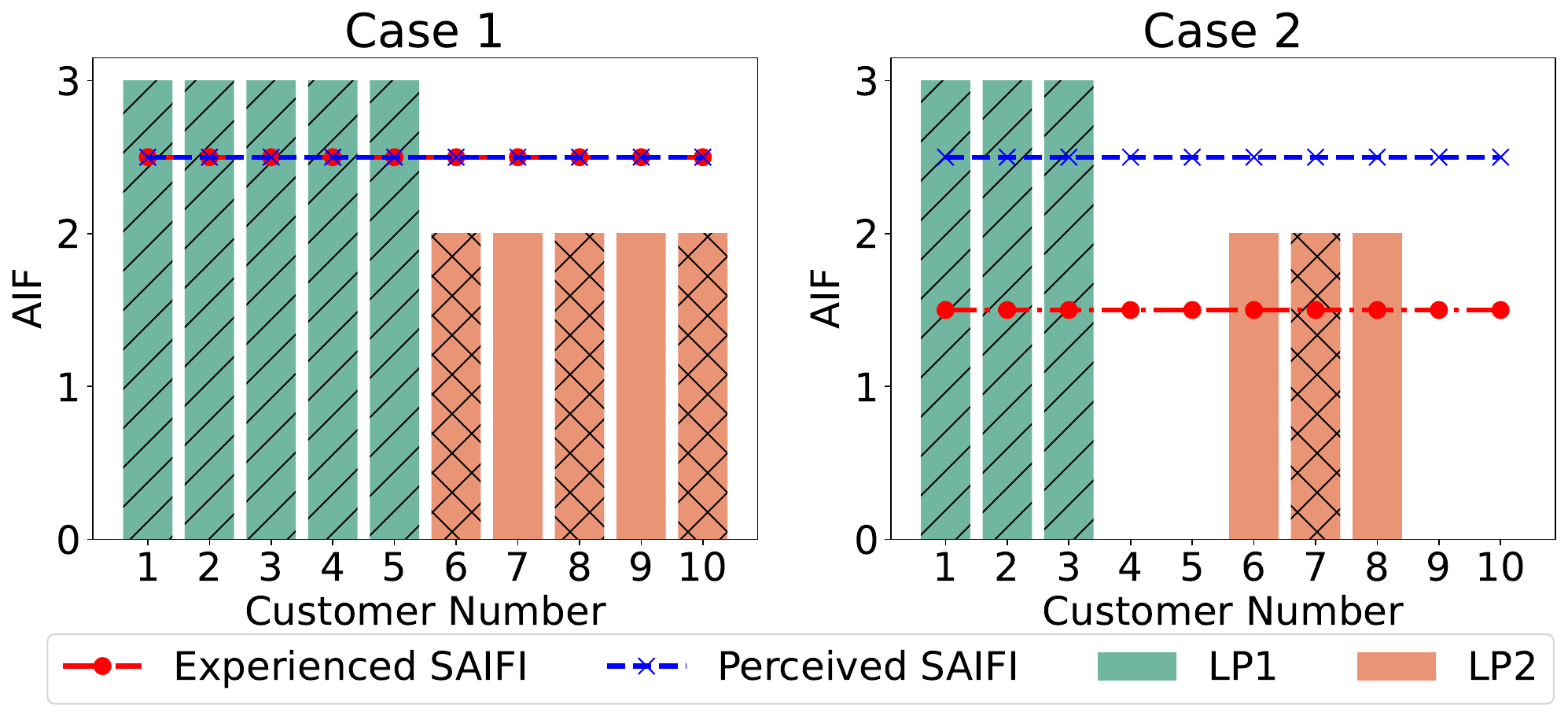}
    \caption{Case 1: no DER; Case 2: DER at \{4,5,9,10\}.}
    \label{fig:percvsexp}
\end{figure}

\textbf{Case 1:} Without DERs, perceived and experienced reliability coincide; each customer is equally exposed to interruptions.  

\textbf{Case 2:} When 40\% of customers adopt BTM DERs (customers 4, 5, 9, and 10), those customers no longer experience interruptions. As shown in Fig.~\ref{fig:percvsexp}, the indices calculated from the DSO’s perspective differ significantly from those experienced by customers. This example highlights that load-point--based methods cannot quantify the reliability value provided by BTM DERs and motivates the need for an approach that captures customer-level outcomes.  

\textbf{Motivation:}  
As residential DER adoption accelerates, reliability assessment must move beyond average indices reported from the utility perspective. Modern customers expect not only continuity of supply but also resilience, flexibility, and recognition of their DER contributions. Ignoring heterogeneity in adoption risks undervaluing customer investments and misguiding planning decisions. Predictive methods must therefore integrate probabilistic adoption models and account for PV and ES component reliability, enabling fair and forward-looking reliability evaluation.  

\textbf{Scope of this work.}  
This paper proposes a modular, bottom-up reliability assessment framework that augments conventional methods. 
The framework models residential systems with configurable BTM DER adoption, integrates PV and ES component reliability, and applies adaptive Monte Carlo simulation to estimate not only mean indices but also the distribution of customer-level outcomes. 
The approach is demonstrated on a modified RBTS Bus 4 system with 16 joint PV--ES adoption scenarios.  

\textbf{Contributions.}  
The contributions of this paper are summarized as follows:
\begin{itemize}
    \item We propose a bottom-up probabilistic framework that evaluates distribution system reliability by aggregating residential-level DER adoption into system indices, in contrast with conventional top-down approaches that compute system indices directly from load-point averages reported at the utility level.
    \item We introduce a novel adoption-modeling approach that uses probability distributions to represent micro-level DER penetration.  
    \item We incorporate PV and ES component reliability into system-level reliability indices.  
    \item We employ an adaptive Monte Carlo simulation method to estimate not only SAIFI/SAIDI means but also their distributional variability across customers.  
    \item We demonstrate the framework on a modified RBTS Bus 4 case study, analyzing 16 joint PV--ES adoption scenarios.  
\end{itemize}

The remainder of this paper is organized as follows: Section~2 describes the residential system reliability evaluation. Section~3 presents the probabilistic formulation. Section~4 introduces the DER penetration models. Section~5 details the Monte Carlo simulation method and its application to the RBTS system.

\section{Residential system Reliability evaluation}
To estimate the reliability of a distribution system with residential DER, it is imperative to understand the reliability of the residential system and the factors affecting it.  Quantifiable metrics for the residential system reliability evaluation are Average interruption frequency (AIF), Average interruption duration (AID), and Average energy not served (AENS).
\subsection{Residential system model}
\begin{figure}
    \centering
    \includegraphics[width = \linewidth]{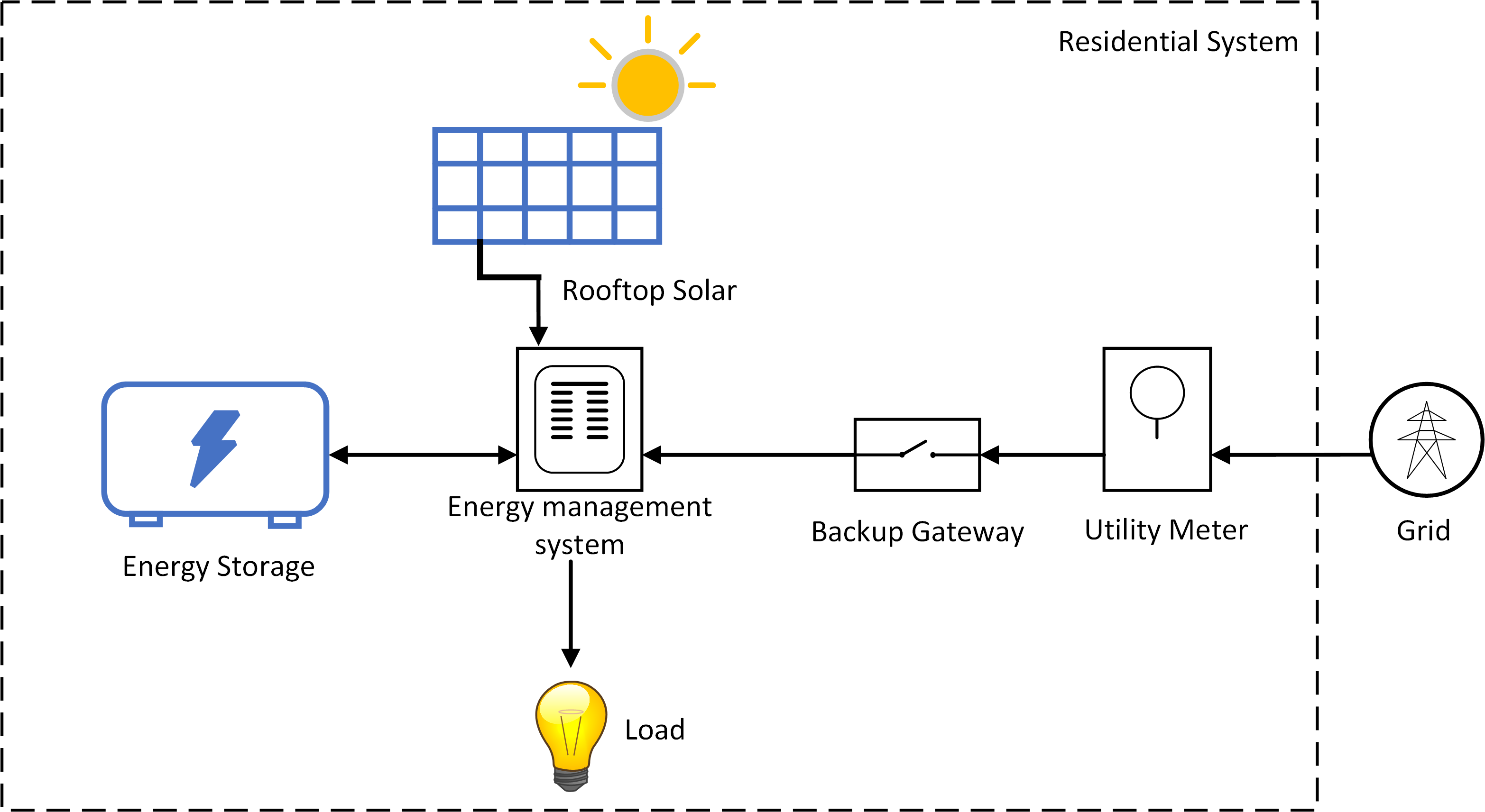}
    \caption{General-purpose residential system configuration.}
    \label{fig:Res_system}
\end{figure}

In this paper, we define a general purpose configuration of a residential system. This configuration has three main components: the load, the DER system, and the Energy management system (EMS). The EMS controls the DER operations and connection to the grid. The EMS operates the residential system primarily in self-consumption mode, where the DER system acts as the primary supply and the utility connection acts as a backup to meet demand not served by the DER. The popular choice of DERs for residential systems is the rooftop PV and ES, and the general purpose configuration reflects both rooftop PV and ES, as shown in the figure \ref{fig:Res_system}. The general purpose model can be reduced to specific configurations based on the DER employed, such as PV only, ES only, or a no DER case where the configuration reduces to a regular residential system. 
\subsubsection{PV system model}
The output of the PV system at any time-step $t$ is modeled as follows, 

\begin{equation}
    \mathrm{PV}_{\text{out}}(t) = \mathrm{PV}_{\text{cap}} \cdot d \cdot \mathrm{GHI}(t)
\end{equation}
where, $\mathrm{PV}_{\text{out}}$ is the output from the rooftop PV system, $\mathrm{PV}_{\text{cap}}$ is the nameplate capacity, $d$ is the derating factor and $\mathrm{GHI}$ is the global horizontal irradiation at the location of the PV system.

%%%%%%%%%%%%%%%%%%%%%%%%%%%%%%%%%%%%%%%%%%%%%%%%%%%%%%%%%%%%%%%%%%%%%%
%%%%%%%%%%%%%%%%%%%%%%%%%%%%%%%%%%%%%%%%%%%%%%%%%%%%%%%%%%%%%%%%%%%%%%

\subsubsection{ES system model}
ES is modeled using the State of Charge (SOC) of the system where SOC is percentage that represents the fraction of the maximum capacity.

\begin{gather}
        \mathrm{SOC}(t) = \mathrm{SOC}(t-1) + \frac{\left( \eta_c Ch(t) - \frac{D(t)}{\eta_d} \right)}{\mathrm{ES}_{\text{cap}}}\\
        \mathrm{ES}(t) = \mathrm{SOC}(t) \cdot \mathrm{ES}_{\text{cap}}\\
        0 \leq Ch(t) \leq Ch_{\text{max}}\\
        0 \leq D(t) \leq D_{\text{max}}\\
        0 \leq \mathrm{SOC}(t) \leq \mathrm{SOC}_{\text{max}}
\end{gather}
where, $\eta_c$ and $\eta_d$ are the charging and discharging efficiency respectively, $Ch$ and $D$ are the charging and discharging energy at a given time-step and $\mathrm{ES}$ is the available energy within the storage at a given time-step.

\subsubsection{Residential system energy balance}
The energy balance for the residential system is modeled as follows:

\begin{equation}\label{eq:energy_balance}
    l_n(t) = l(t) - \mathrm{PV}_{\text{out}}(t) \pm \mathrm{ES}(t)
\end{equation}
where, $l(t)$, represents the load of the residential system and $l_n(t)$ represents the net load of the residential system at time $t$. Net load is the load seen by the utility at the residential system utility meter. 

\subsection{Residential system reliability metrics}

In the context of the residential system model presented, these reliability indies are evaluated as follows: Net load determines the state of the energy balance within the residential system and it has three states. \textbf{Excess energy}: When the PV produces more energy than the residential system can consume or store. \textbf{Unserved load}: When the PV output and the ES combined cannot meet the residential system consumption. \textbf{Balanced load}: When the PV output and the ES system operation perfectly meet the residential system consumption. 

% \begin{itemize}
%     \item \textbf{Excess energy}: When the PV produces more energy than the residential system can consume or store.
%     \item  \textbf{Unserved load}: When the PV output and the ES combined cannot meet the residential system consumption.
%     \item \textbf{balanced load}: When the PV output and the ES system operation perfectly meet the residential system consumption. 
% \end{itemize}

To put this into perspective, form the energy balance equation in (\ref{eq:energy_balance}), if $l_n(t)$ is positive, it means there is an unserved load and energy is drawn from the distribution system. If  $l_n(t)$ is negative, it indicates the availability of excess energy which can be fed back to the distribution system or curtailed, and zero indicates that the onsite energy generation meets the load. 

We define a binary state variable to represent the state of the residential system energy balance as follows:

\begin{equation}\label{eq:state_energy_balance}
    s_{\text{n}}(t) = 
\begin{cases} 
    0 &  l_n(t) > 0 \\
    1 &  l_n(t) \leq 0 
\end{cases} 
\end{equation}

The reliability metrics not just depend on the state of the residential system energy balance but also depend on the state of the load point to which the residential system is connected. The state of the load point depends on the reliability characteristics of the load point. The load point reliability characteristics are usually defined by the load point reliability metrics evaluated using the current methods and the synthetic sequential load point history for a given time horizon can be generated by sequentially calculating the up and down times of the load point as follows:

\begin{gather}
        \mathrm{T}_{\text{TTF}} = -\frac{\ln(\mathrm{R}_1)}{\lambda_{\text{lp}}}\\
            \mathrm{T}_{\text{TTR}} =  -\frac{\ln(\mathrm{R}_2)}{\mu_{\text{lp}}}
\end{gather}

where, $\mathrm{T}_{\text{TTF}}$ is the time to failure of the load point, $\mathrm{T}_{\text{TTR}}$ is the time to repair of the load point, $\lambda_{\text{lp}}$ is the failure rate, and $\mu_{\text{lp}}$ is the repair rate of the load point and $\mathrm{R}$ is a random number drawn from a uniform distribution  . We define a binary state variable to describe the state of the load point at time t as follows:

\begin{equation}\label{eq:state_lp}
    s_{\text{lp}}(t) = 
\begin{cases} 
    0 & \text{if LP is down} \\
    1 & \text{if LP is up} 
\end{cases} 
\end{equation}

Given the state of the energy balance and the state of the load point, the residential system state can be defined as follows. \textbf{Operating State}: When either the DER meets the local load or the load point is in the operating condition. \textbf{Failed State}: When there is unserved load and the load point is in the failed condition. We define another state variable to describe the state of the residential system as follows:

\begin{equation}\label{eq:state_residential}
    s_r = s_{lp} \lor s_n
\end{equation}

Given the residential system state variable $s_r$, an interruption event occurs when there is a transition from the operating state ($s_r(t-1) = 1$) to the failed state ($s_r(t)= 0$). The logical expression capturing this transition is given as follows:
\begin{equation}\label{eq:Interruption}
    \mathrm{I}(t) = s_r(t-1) \land \neg s_r(t)
\end{equation}

where $\mathrm{I}$ is the Boolean variable representing the interruption event at time $t$. This variable is \textbf{True} for every time $t$, where an interruption starts. The AIF is then calculated by averaging the number of interruption events over the study period as follows:

\begin{equation}\label{eq:AIF}
    \mathbf{AIF} = \frac{\sum_{t=1}^{T} \mathrm{I}}{T}
\end{equation}

The AID is calculated by averaging the duration of interruption over the study period as follows:

\begin{equation}\label{eq:AID}
    \mathbf{AID} = \frac{\sum_{t=1}^{T} s_r}{T}
\end{equation}

% The AENS is calculates in the same way by averaging the total energy not served over the study duration:

% \begin{equation}
%     \mathbf{AENS} = \frac{\sum_{t=1}^{T} s_r * l(t)}{T}
% \end{equation}
We used a Monte Carlo simulation  method to estimate the reliability indices of the residential system \cite{karngala2023impact}.

\section{Probabilistic approach to system reliability assessment}
This section provides a detailed description of the bottom up probabilistic approach to evaluate the reliability of the distribution system using the general purpose residential system configuration as a basic building block. 
In the context of system reliability, the distribution system is viewed as a set of load points and the associated customers connected to the load point.

We define the set of load points as $\mathcal{L}$, where 
\begin{equation}
    \mathcal{L} = \{L_1, L_2, \dots, L_n\}
\end{equation}
Each $L_i$ represents a specific load point for $i = 1,2,\dots, n$. For each load point $L_i$, there is an associated set of customers $C_i$ connected to it. 
\begin{equation}
    C_i = \{c_{i1}, c_{i2}, \dots, c_{im_i}\}
\end{equation}
where $m_i$ is the number of customers connected to the load point $L_i$ and each $c_{ij}$ represents an individual customer connected to $L_i$.
To represent all the sets of customers connected to the load points, we define 
\begin{equation}
    \mathcal{C} = \{C_1, C_2,\dots, C_n\}
\end{equation}
where $C_i$ is the set of customers connected to the corresponding load point $L_i$. The relation ship between the set of load points and the set of customers is captured as follows:
\begin{equation}
    \forall L_i \in \mathcal{L}, \exists C_i \in \mathcal{C}
\end{equation}

Residential system reliability metrics quantify the susceptibility to interruptions and are influenced by several parameters, namely $l$, $\mathrm{PV}_{cap}$, $\mathrm{ES}_{cap}$, and the reliability of the load point which is quantified by the state of the load point $s_\text{lp}$. Together, these variables articulate the circumstances under which the residential system may experience failures, thereby contributing to the residential system reliability metrics. Using the defined notation, we can express the reliability metrics as a function of the influencing variables as follows.

\begin{equation}\label{eq:AIF_fn}
    \mathbf{AIF}_{c_{ij}} = g\left( \mathrm{PV}_{c_{ij}, \text{cap}}, \; \mathrm{ES}_{c_{ij}, \text{cap}}, \; l_{c_{ij}}, \; s_{L_i, \text{lp}} \right)
\end{equation}
\begin{equation}
    \mathbf{AID}_{c_{ij}} = h\left( \mathrm{PV}_{c_{ij}, \text{cap}}, \; \mathrm{ES}_{c_{ij}, \text{cap}}, \; l_{c_{ij}}, \; s_{L_i, \text{lp}} \right)
\end{equation}
\begin{equation*}
    \forall c_{ij} \in \mathcal{C}
\end{equation*}

Here $g(\cdot)$ and $h(\cdot)$ denote the reliability mapping functions introduced in the Monte Carlo formulation. The function $g(\cdot)$ maps a realization of adoption ratios $(X,Y)$ and outage conditions to the interruption frequency contribution (AIF) for a representative customer, i.e., the number of interruptions experienced during that trial. Similarly, $h(\cdot)$ maps the same inputs to the interruption duration contribution (AID), i.e., the number of outage hours incurred by the customer.

For the purpose of illustration, only $\mathbf{AIF}$, is used to show the bottom up formulation to evaluate system metrics. 
At this point, we introduce different layers of abstraction based on the granularity of the analysis. 

\textbf{Abstraction layer 1:} In the first layer, the reliability provided by the load points is uniform across the distribution system. In the context of predictive reliability assessment, it is a reasonably fair abstraction given the goal is to quantify the BTM residential DER contribution to the distribution system reliability. Therefore, $s_{lp}$ has a consistent influence on the $\mathbf{AIF}_{c_{ij}}$, allowing us to simplify the function to:

\begin{equation}
\mathbf{AIF}_{c_{ij}} = g\left( \mathrm{PV}_{c_{ij}, \text{cap}}, \mathrm{ES}_{c_{ij}, \text{cap}},  l_{c_{ij}} \right)
\end{equation}

To address the disparity in load magnitude, PV system sizes, and ES capacities among different residential systems, we normalize these variables with the peak load ($l_{c,p}$), yielding:

\begin{equation}
    \begin{split}        
    \mathbf{AIF}_{c_{ij}} & = g\left( \frac{\mathrm{PV}_{c_{ij},\text{cap}}}{l_{c_{ij},p}}, \; \frac{\mathrm{ES}_{c_{ij},\text{cap}}}{l_{c_{ij},p}}, \; \frac{l_{c_{ij}}}{l_{c_{ij},p}} \right) \\
    & = g\left( \mathrm{PV'}, \; \mathrm{ES'}, \; l' \right)
    \end{split}
\end{equation}

The above equation means that for any residential unit connected to the distribution system, the $\mathbf{AIF}$ is a function of the load profile $l'$ and the normalized PV and ES system sizes. Assuming homogeneity in load profiles across residential systems, with variations in load values being primarily due to the magnitude of the load, $l'$ is consistent for all residential systems, leading to a further simplified model:

\begin{equation}\label{eq:AIF_X_Y}
    \mathbf{AIF}_{c_{ij}} = g\left( \mathrm{PV'}, \; \mathrm{ES'} \right) = g\left( X, \; Y \right)
\end{equation}
for simplicity we define $\mathrm{PV'}= X$ and $\mathrm{ES'}= Y$, where  $X$ and $Y$ as continuous random variables with a joint density $f_{XY}(x,y)$, representing the penetration or forecasted penetration of DERs among the residential systems connected in the distribution system. The \textbf{SAIFI} is simply the expected value of the $\mathbf{AIF}$ over the joint density of $f_{XY}(x,y)$

\begin{equation}
\begin{split}
    \mathbf{SAIFI} & = \mathop{\mathbb{E}}\left( \mathbf{AIF}_{c_{ij}} \right)\\
    & = \mathbb{E}\left( g(X, \; Y) \right)
\end{split}
\end{equation}
where, the expectation of $g(X,Y)$ with respect to the joint density is given by

\begin{equation}
    \mathbb{E}\left[g(X, Y)\right] = \iint g(x, y) f_{XY}(x, y) \, \mathrm{d}x \, \mathrm{d}y
\end{equation}

\textbf{Abstraction layer 2:} In the second layer of abstraction, the reliability provided by the load points is non uniform across the distribution system. Therefore (\ref{eq:AIF_fn}) is simplified to the following: 

\begin{equation}
\mathbf{AIF}_{L_i} = g\left( \mathrm{PV}_{c_{ij}, \text{cap}}, \mathrm{ES}_{c_{ij}, \text{cap}},  l_{c_{ij}} \right)
\end{equation}
\begin{equation*}
    \forall c_{ij} \in C_i
\end{equation*}
where,  $s_{lp}$ has a consistent influence on the residential systems connected to the load point $L_i$. 
Further simplification as in the earlier abstraction layer will lead to 

\begin{equation}
    \mathbf{AIF}_{L_i} = g\left( \mathrm{PV'}, \; \mathrm{ES'} \right) = g\left( X, \; Y \right)
\end{equation}
in this case the joint density $f_{L_i,{XY}}(x,y)$ represents the penetration or forecasted penetration of DERs across the residential systems connected to the load point $L_i$.

We define the load point average interruption frequency index for a load point $L_i$ as the expected value of $\mathbf{AIF}$ over the residential systems connected to $L_i$. It simply is the \textbf{SAIFI} for the load point $L_i$

\begin{equation}
\begin{split}
    \mathbf{LAIFI}_{L_i} & = \mathop{\mathbb{E}}\left( \mathbf{AIF}_{L_i} \right)\\
    & =\iint g(x, y) f_{L_i,{XY}}(x,y) \, \mathrm{d}x \, \mathrm{d}y
\end{split}
\end{equation}

The model for the load point is then extended to a distribution system with multiple load points each with a different expectation of the \textbf{AIF}.  System Average Interruption Frequency Index is obtained by taking a weighted average of each load point's average interruption frequency. 

\begin{gather}\label{eq:layer1}
        \mathbf{SAIFI} = \sum_{i=1}^n w_{L_i} * \mathbf{LAIFI}_{L_i}\\
            w_{L_i} = \frac{|{C_i}|}{\sum_{i=1}^n |C_i|}
\end{gather}

Similar to these two abstraction layers, another layer that could be of practical importance is at the feeder level. In this layer, the reliability provided by the load points is uniform across a distribution feeder. 

In all the layers of abstraction, we are primarily quantifying the reliability using an average metric such as $\mathbf{SAIFI}$. While the average provides an excellent basis to quantify the contribution of DERs to system reliability, it would be more insightful and valuable to understand the underlying distribution of the residential reliability metric for a given distribution of DER penetration. From (\ref{eq:AIF_X_Y}) we can see that the random variable $\mathbf{AIF}$ is a transformation of the random variables $X$ and $Y$ through the function $g$. To determine the probability distribution of $\mathbf{AIF}$, we can leverage the joint PDF $f_{XY}(x,y)$. Since an explicit form of $g$ is not available, numerical methods such as Monte Carlo Simulations combined with kernel density estimation can be employed to approximate the desired PDF of the reliability metric. 

%%%%%%%%%%%%%%%%%%%%%%%%%%%%%%%%%%%%%%%%%%%%%%%%%%%%%%%%%%%%%%%%%%%%%%%%%%%%%%%%%%%%%%%%%%%%%%%%%%%%%%%%%%

\section{Modeling DER penetration}
In the proposed probabilistic model, the necessary information to evaluate system reliability is the joint density $f_{XY}(x,y)$. This density refers to the penetration of edge level DERs at the residential level. The advantages of such modeling, and the technical details are presented in the following subsections.
\subsection{Macro vs Micro Perspectives}
DER penetration is usually quantified with aggregates at the bulk system level or stratified into different customer sectors at the distribution system level. It is defined as the total installed capacity of the DER across a system, relative to the peak load of that system.

In the context of PV and ES, they are defined as follows:
\begin{gather}
    \phi_{\text{PV}} = \frac{\text{Total Installed PV capacity (kW)}}{\text{Total peak load(kW)}} \\
    \phi_{\text{ES}} = \frac{\text{Total Installed ES capacity (kWh)}}{\text{Total peak load(kW)}}
\end{gather}
These macro penetration metrics $\phi_{\text{PV}}$, $\phi_{\text{ES}}$ provide a holistic view of how saturated a given system or region is with the adopted DER technology. Although useful, this high-level perspective can obscure variations and patterns at the residential levels. In this rapidly evolving landscape of edge-level DERs, understanding the penetration of DER can no longer remain a macro-problem. Instead, capturing the intricacies of individual behaviors, choices, and motivations is paramount as the actual dynamics of the energy transition plays out at this granular edge level. By modeling the penetration at the microlevel, one can peel back the layers of aggregate statistics to reveal the vibrant mosaic of individual decisions and factors that drive the bigger picture.

% \textbf{Contextual example:} Consider a dataset of 100 households within a region. A macro-level observation might reveal a total daily solar generation of 1000 kWh. However, this aggregated data doesn't specify distribution—whether it’s homogeneous across households or if certain households significantly contribute more than others. A micro-level analysis provides clarity on these distributions.

\subsection{Quantitative Modeling of Micro-level Penetration}

The primary goal is to model the micro-level penetration of DER installations at individual residential units within a  distribution system. By characterizing this penetration as a probabilistic phenomenon, we aim to make informed predictions, simulate scenarios, and derive insights about the distribution system's reliability.

Let $X$ and $Y$ represent the PV and ES penetration of a randomly chosen residential system from a defined distribution system, where

\begin{gather}
    X= \frac{\mathrm{PV}_{\text{cap}}}{l_\text{p}}\\
    Y = \frac{\mathrm{ES}_{\text{cap}}}{l_\text{p}}
\end{gather}
$X$ indicates how the rooftop PV system capacity compares to the residential system peak electricity demand. 

\begin{itemize}
    \item \textbf{$X = 1$:} PV system capacity matches the peak load
    \item \textbf{$X > 1$:} PV system has excess capacity relative to the peak load
    \item \textbf{$X < 1$:} PV system might not fully cover the peak load
\end{itemize}
$Y$ indicates how much energy the ES system can hold relative to the residential system peak electricity demand. 

\begin{itemize}
    \item \textbf{$Y = 1$:} ES system can support the residential system peak load for an hour.
    \item \textbf{$Y > 1$:} ES system can support the residential system peak load for more than an hour. 
    \item \textbf{$Y < 1$:} ES system cannot supply the residential system peak load for a full hour. 
\end{itemize}
Together, these ratios provide insight into how self-sufficient a residential system might be in terms of generating and storing its own electricity. 

We model the random variables X and Y as probability distributions to characterize the variations and patterns in DER installations among distinct residential units within a distribution system. The probability distribution parameters will describe specific aspects of the micro-level DER penetration within the distribution system.

%  To illustrate, consider the micro level penetration of residential PV, assuming that $X$ follows a particular distribution 

% \begin{itemize}
%     \item \textbf{Central tendency}: (e.g., Mean, Median) Describes the average or typical rooftop PV penetration rate among residential systems. A higher mean implies a greater trend towards higher individual PV adoption.
%     \item \textbf{Spread}:  (e.g., Variance, Standard Deviation) Indicates the variability in PV installations. A larger spread suggests diverse behaviors in PV adoption among households.
%     \item \textbf{Skewness}: A measure of the distributions asymmetry. Positive skewness indicates that many residential systems might have above-average PV installation. 
%     \item \textbf{Kurtosis}: Describes the tailedness of the distribution. High kurtosis might indicate that many households either have very high or very low PV installations relative to peak demand. 
%     \item \textbf{Bounds}: The range or limits of the distribution. In the given context, the minimum installed PV size and the maximum installed PV size, relative to their respective peak loads. 
% \end{itemize}

% \subsection{Joint modeling of X and Y}
Examining the marginal distributions of $X$ and $Y$ yields insights into individual characteristics; however, the intricate dynamics and the co-dependence between rooftop PV and ES systems necessitate joint modeling to understand their combined implications on the distribution system.
% We can derive a joint probability distribution by examining $X$ and $Y$ in tandem, explaining the likelihood of concurrent adoption patterns and their compounded effects on the grid. The interdependence between $X$ and $Y$ propels the necessity of understanding their correlation.
In this work, we construct scenarios for the joint adoption of $X$ and $Y$ and assign specific correlations, allowing us to explore various ways these variables might interact.

\subsection{Marginal Distributions for X and Y}
\label{sec:adoption}
Field penetration data provides a solid foundation for accurately modeling DER penetration. However, a significant challenge arises due to the shallow adoption of BTM DER, especially in ES in the current energy landscape. This shallow adoption hinders our ability to construct meaningful, data-driven adoption scenarios. Consequently, we rely heavily on hypothetical constructions and theoretical models. By shaping distributions with these theoretical methods, we can approximate the specific patterns that might emerge in reality. These hypothetical models are tools for current understanding and are invaluable in predicting future Photovoltaic (PV) penetration scenarios. By integrating forecasts and considering potential shifts in policy and technology, we can use these theoretical constructs to anticipate and prepare for changes in the energy landscape despite the current limitations in real-world data. It is worth noting, however, that the core objective of this paper diverges slightly. Rather than focusing on the details of creating these distributions, our primary interest lies in developing a framework to utilize these distributions to derive scenario-based insights on system reliability.

We define four different adoption types each for both PV and ES micro-level adoption. The adoption types are Limited (L), varied (V), median-focused (MF), and higly-concentrated (HC). Before discussing the choice of the distributions used to model the random variables, the range of the random variables needs to be defined. As of 2021, the average size of residential solar installations in the U.S. was approximately 7kW \cite{barbose2021behind}, compared to the average peak load of 4kW for a typical home, yielding a ratio of 1.75. This average ratio, indicative of a significant margin over immediate energy needs, also sets a reference for understanding the variability in solar energy production. By defining a range for X from 0 to 3.5, we accommodate the diversity of household energy profiles, acknowledging that some homes may not have solar capacity (hence the lower bound of  0). In contrast, others might have installations capable of generating up to twice the average output, providing a flexible scope that captures the spectrum of potential homeowner experiences.

Similarly, with residential ES, the popular residential ES solutions have capacities of 13.5kWh \cite{barbose2021behind}. Considering this system as an average, it can support the peak load of an average household for approximately 3.375 hours. By extending the range from 0 to 6.75 hours, we encompass the full breadth of storage capacities, considering that some homes may opt not to have ES. In contrast, others may install systems that offer double the average duration of peak load coverage, thus ensuring preparedness for scenarios from short-term outages to extended periods without grid power.  

These ranges are not arbitrary but thoughtfully chosen to represent the variability inherent in residential energy solutions. The average values give us a baseline, while the extended ranges ensure we capture the real-world fluctuations, both in terms of solar generation potential and energy storage capacity, thus providing a more comprehensive picture that homeowners can relate to and plan for. 

\begin{figure}
    \centering
    \includegraphics[width = \linewidth]{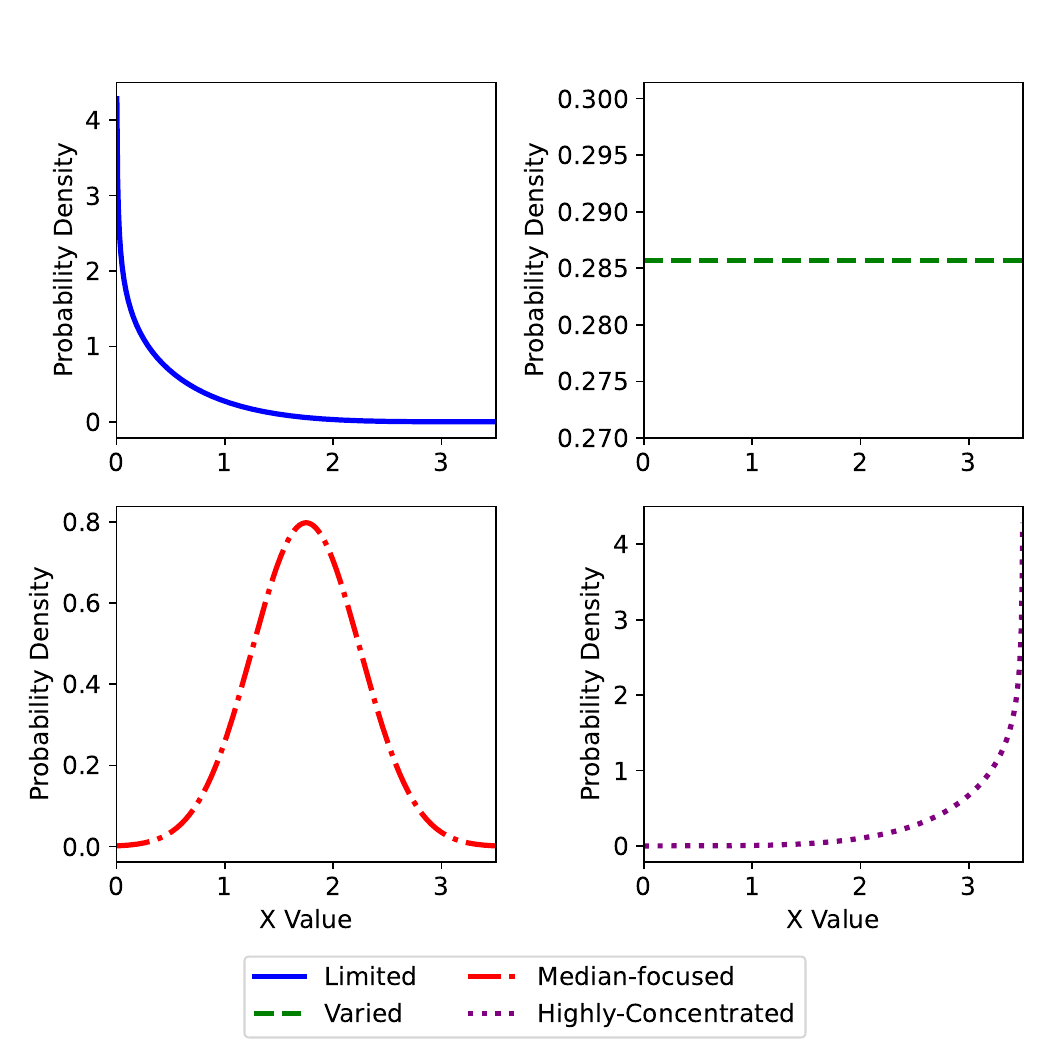}
    \caption{Marginal adoption scenarios for X.}
    \label{fig:marginals}
\end{figure}

The choice of distribution and the adoption type for both X and Y are described as follows and are visualized in Fig.  \ref{fig:marginals}.

\noindent\textbf{Limited adoption:} This adoption type is modeled using a scaled Beta distribution as follows.
\begin{gather}\label{beta}
         f(x'; \alpha, \beta) = \frac{1}{scale}\cdot\frac{x'^{\alpha - 1}(1 - x')^{\beta - 1}}{B(\alpha, \beta)} \\
         x' = \frac{x-loc}{scale}
\end{gather}
where $\alpha = 0.75$ and $\beta = 5$. The distribution is skewed towards the lower end of the range and the support is shifted from the standard [0,1] interval to [0, 3.5] for X and [0, 6.75] for Y by the $loc$ and $scale$ parameters. 
    \begin{itemize}
        \item X: This scenario assumes a conservative approach to the adoption of PV systems. Only a tiny fraction of homes have PV installations. Among those that do, the PV capacity is closer to the lower end of the defined range for $X$. In essence, both the number of houses with PV systems and the sizes of these systems are limited.
        \item Y:ES systems are not widely adopted. Among those with them, the storage capacity is minimal, possibly due to high costs, lack of awareness, or limited perceived need. 
    \end{itemize}
    
\noindent\textbf{Varied adoption:} This adoption type is modeled using a uniform distribution. 
\begin{equation}
    f(x;a,b) = \begin{cases}
          \frac{1}{b-a} & \text{for } a \leq x \leq b \\
   0 & \text{otherwise}
    \end{cases}
\end{equation}
where $a$, and $b$ are the range limits for the random variables.
    \begin{itemize}
        \item X:  Here, while the number of residential systems adopting PV might be more, the size of the installations varies greatly. Some houses might have a minimal PV capacity, while others approach the upper limit.
        \item Y: Many homes have adopted ES systems, but their installed capacities vary greatly. While some homes might have minimal storage, others might approach the upper end of the capacity spectrum.
    \end{itemize}

\noindent\textbf{Median-focused adoption:} A truncated normal distribution is used to model the median-focused adoption. 
\begin{equation}
      f(x; \mu, \sigma, a, b) = \frac{\phi\left(\frac{x - \mu}{\sigma}\right)}{\sigma \left(\Phi\left(\frac{b - \mu}{\sigma}\right) - \Phi\left(\frac{a - \mu}{\sigma}\right)\right)}
\end{equation}
for $a<x<b$, with $\phi$ and $\Phi$ denoting the PDF and the CDF of the standard normal distribution respectively. The middle value of the range ($\frac{a+b}{2}$) is taken as the mean ($\mu$) of the distribution with a relatively small standard deviation ($\sigma$). For X, $\sigma = 0.5$ and for Y, $\sigma=1$.
    \begin{itemize}
        \item X: The majority of homes with PV installations have capacities around a median value. This could be due to the standardization of the popular packages offered by installation companies.
        \item Y: Most homes with ES systems have capacities close to the median value. This could be due to specific popular product offerings or household ES needs.
    \end{itemize}
    
\noindent\textbf{Highly concentrated:} Beta distribution as defined in (\ref{beta}) is used to model this adoption scenario, where $\alpha = 5$ and $\beta = 0.75$. The shape parameters are modified to reflect high concentration near the upper bound of the range. 
    \begin{itemize}
        \item X: Here, most residential systems that adopt PV go for the higher end of the capacity spectrum. These are the homes maximizing their solar potential, possibly due to higher incentives, more significant energy needs, or other drivers.
        \item Y: In this scenario, homes that adopt ES systems tend to install capacities on the higher end of the spectrum. Reasons could include larger homes, areas with unreliable grid power, or incentives that make larger systems more attractive.
    \end{itemize}

\subsection{Joint adoption scenarios}

\begin{table}
\caption{Correlation for Joint Scenarios}
\label{tab:adoption scenario table}
\centering
\begin{tabular}{ccccc}
\hline
\diagbox[dir=NW]{\textbf{X}}{\textbf{Y}} & \textbf{Limited} & \textbf{Varied} & \textbf{Median} & \textbf{Highly Conc.} \\
\hline
\textbf{Limited} & 0.7 & 0.2 & 0.4 & -0.2 \\
\hline
\textbf{Varied} & 0.2 & 0.4 & 0.3 & 0.1 \\
\hline
\textbf{Median} & 0.3 & 0.4 & 0.5 & 0.3 \\
\hline
\textbf{Highly Conc.} & 0.2 & 0.2 & 0.3 & 0.8 \\
\hline
\end{tabular}
\end{table}

Given these marginal distributions, we construct joint adoption scenarios from these individual marginals.  Table \ref{tab:adoption scenario table} provides a structured representation of hypothetical pairwise correlations between adoption scenarios for two variables, X and Y. \textbf{Rows} represents different adoption scenarios for Variable X. \textbf{Columns} represents different adoption scenarios for Variable Y. \textbf{Values within the table} indicate the assigned correlation coefficients between the corresponding scenarios of X and Y. It is essential to note that this table is not a correlation matrix. While a correlation matrix represents correlations between multiple variables, this table showcases the relationship between different scenarios of two specific variables. In essence, the table offers insights into how different adoption scenarios for X might relate to or influence corresponding scenarios for Y. It serves as a guide for understanding potential relationships between these variables under different circumstances.

% While correlation provides a scale-invariant measure, covariance reveals the raw metric of joint variability. We can compute the covariance matrix once we know the marginal distributions of $X$, $Y$ and the assumed correlation coefficient between them. The details to calculate the covariance matrix are provided in appendix \ref{apen:cov}.

% ============================================================
% CHANGE: Added new subsection 4.E in response to Reviewer 1, Comment 1
% ============================================================

\subsection{Modeling Assumptions and Applicability}
\label{sec:assumptions}

To clarify the scope of the proposed framework, we summarize below the key assumptions adopted in this work and discuss their applicability across different distribution system contexts.

\textbf{Uniform Load Point Reliability:} In the first abstraction layer, it is assumed that the reliability indices are uniform at all load points. This assumption is appropriate for urban and suburban systems, where infrastructure quality and operational standards are relatively homogeneous. However, in rural networks, the length of the feeder and the customer density often result in a heterogeneous reliability performance. The proposed framework is designed to accommodate such nonuniform load point indices through a second abstraction layer. In this paper, we focus on the uniform case to illustrate the methodology; extending the analysis to heterogeneous load point reliability will be pursued in future work.  

 \textbf{Homogeneity of Normalized Load Profiles:} Residential load shapes are assumed to be homogeneous once normalized by peak demand. Empirical datasets such as NREL’s ResStock~\cite{ResStock} confirm that load profiles within a climatic and socio-economic region exhibit consistent temporal patterns, with variation largely explained by magnitude rather than shape.  

\textbf{PV and ES Component Reliability:} The framework also accounts for the reliability of behind-the-meter DER components. Rooftop PV is modeled through AC modules (PV panel plus micro-inverter), and storage through inverter-integrated battery modules, each subject to independent failure and repair processes. Including these subsystem dynamics ensures that customer-level reliability indices reflect not only grid-side interruptions but also the availability of the DER hardware itself. This treatment extends the framework beyond adoption patterns alone and aligns with prior component-level reliability models established in our earlier work~\cite{karngala2023impact}.

\textbf{Adoption Distributions:} Marginal adoption distributions for PV and ES are modeled using theoretical forms (Beta, Uniform, Truncated Normal) to represent different adoption archetypes. This scenario-based approach is consistent with prior distributed energy studies~\cite{barbose2021behind, kiray2025scenario, baswaimi2024planning}. The objective is not to forecast specific adoption trajectories, but to illustrate how the framework evaluates system reliability under diverse but plausible penetration patterns.  

 \textbf{PV--ES Correlation:} Correlation coefficients in the joint adoption scenarios (Table~\ref{tab:adoption scenario table}) are selected to span a plausible range of interactions. While detailed empirical correlation data remain limited, emerging evidence suggests strong complementarity. For example, the NREL LA100 study~\cite{nrel_la100} projects 1.1--1.5~GW of distributed storage co-adopted with 3--4~GW of rooftop PV in Los Angeles by 2045, corresponding to roughly 30--40\% of PV capacity being paired with storage. In the near term, EIA data~\cite{eia2024_storage} indicate that more than 50\% of new residential PV installations in California in 2024 were paired with storage. Together, these findings motivate the inclusion of moderate-to-high positive correlations (0.3--0.8) in the scenario matrix. At the same time, survey and behavioral studies~\cite{bollinger2024nber, fett2021pvstorage, alipour2021bess} show that adoption decisions are also shaped by user perception, economic considerations, and policy incentives, meaning that weaker correlations (0.1--0.3) may also arise in practice. Finally, a small negative correlation (-0.2) is included as a stress-test case to explore conditions where storage adoption may be incentivized independently of PV (e.g., as a resilience investment for backup-only applications). Overall, the correlation scenarios are intended to provide sensitivity bounds rather than precise forecasts, ensuring the robustness of the proposed reliability assessment framework.

 As empirical adoption datasets become available at higher granularity, the proposed framework is designed to directly integrate such data, ensuring that the current scenario-based analysis can evolve into a data-driven validation

\section{Application of the probabilistic approach}
In this section, we apply the probabilistic method to analyze the RBTS test system. We begin by thoroughly explaining the Monte Carlo Simulation framework used to determine the reliability indices and their distributions. Following that, we detail the marginal adoption distributions for PV and ES and the modeling of their joint scenarios. We conclude the section with a presentation and critical analysis of the simulation results.

% ============================================================
% CHANGE: Rewritten Section 5.A with integrated adaptive MC details
% ============================================================

\subsection{Monte Carlo Simulation Framework}
\label{sec:MC_framework}

\begin{figure}
    \centering
    \includegraphics[width = \linewidth]{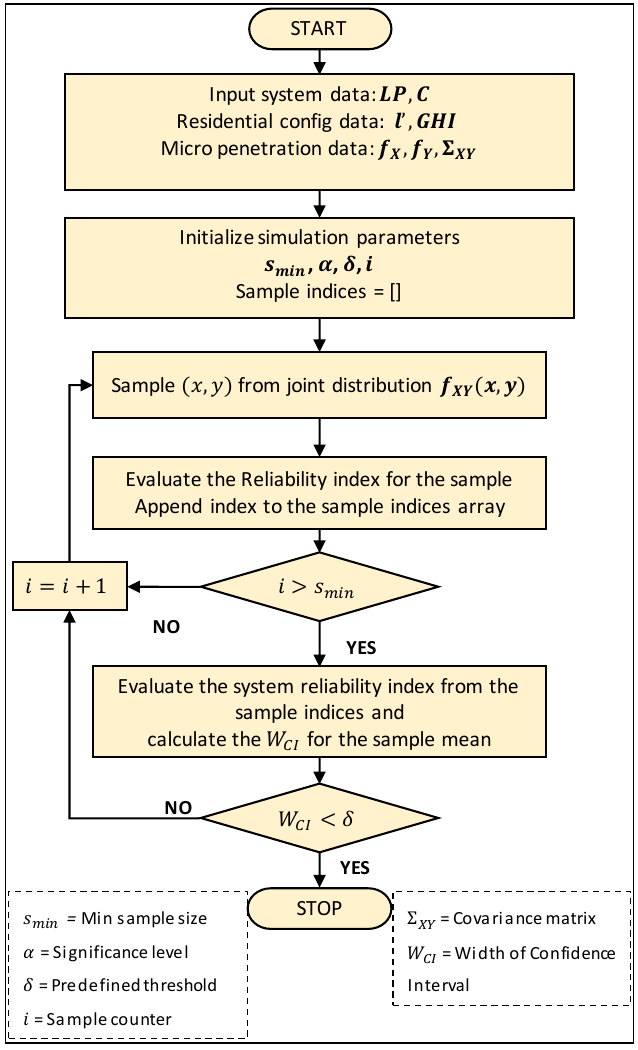}
    \caption{MCS flowchart.}
    \label{fig:flowchart}
\end{figure}

The reliability indices under different DER adoption scenarios are estimated using an adaptive Monte Carlo (MC) simulation framework. The process is structured as a sequence of independent stratified sampling batches, each covering the input probability space with low discrepancy. Within each batch, the system model is evaluated and per-sample reliability metrics (e.g., SAIFI, SAIDI) are accumulated. The overall workflow, shown in Fig.~\ref{fig:flowchart}, iteratively expands the sample size until statistical convergence is achieved. This adaptive strategy avoids the inefficiencies of fixed sample sizes and ensures that precision requirements are met without unnecessary computation.  

\textbf{Estimator updates.} For each reliability metric $k \in \{\text{SAIFI, SAIDI}\}$, running means $\hat{\mu}_k(n)$ and unbiased variances $\hat{\sigma}^2_k(n)$ are updated in streaming form (Welford’s method). This guarantees numerical stability and requires only $O(1)$ memory.  

\textbf{Stopping rule.} Convergence is determined using two criteria:  

\begin{enumerate}
    \item \emph{Minimum information floor:} At least one full batch ($n_{\min} = 10$) must be accumulated before any inference is allowed.  

    \item \emph{Confidence interval precision:} For each metric, the half-width
    \begin{equation}
        h_k(n) = z_{\alpha/2}\,\hat{\sigma}_k(n)/\sqrt{n}
    \end{equation}
    must not exceed a predefined tolerance. In this study, tolerances are set as
    \[
    \varepsilon_{\text{SAIFI}} = 0.005~\text{interruptions/customer-year},
    \]
    \[
    \varepsilon_{\text{SAIDI}} = 0.1~\text{hours/customer-year}.
    \]
    These thresholds correspond to reporting precision at the second decimal place for SAIFI and the first decimal place for SAIDI, consistent with IEEE Std.~1366 reporting practices.   
\end{enumerate}

This criterion ensures that the reported estimates have a reproducible precision level and prevents premature termination due to insufficient sampling.  

\textbf{Parameterization in this study.} We adopt a 95\% confidence level ($\alpha = 0.05$), batch size of 10 (replicate-stratified mode), a maximum sample cap of 2000, and the absolute CI half-width thresholds listed above. Deterministic seeding (42) and per-batch hashing ensure reproducibility, while the final number of samples and achieved half-widths are reported for transparency.  

With these convergence criteria and precision targets established, the following subsection applies the adaptive MC framework to the RBTS Bus 4 system.

\subsection{Application to RBTS Bus 4}
\textbf{System setup and abstraction:}
The RBTS distribution test system consists of seven feeders, 38 load points, and 4779 customers \cite{allan1991reliability}. 
For this study, feeders 2, 5, and 6, which supply nine industrial customers are removed, leaving approximately 4700 residential customers served by 22 load points. 
Because the analysis in this paper is conducted at the \emph{first abstraction layer}, all load points are assigned uniform reliability characteristics. 
To enforce this, the baseline system indices of the modified RBTS (excluding industrial feeders) are computed and applied as the basic load point indices across the network. 
This homogenization ensures that the computational effort depends primarily on the number of Monte Carlo samples rather than the system size, enabling the framework to scale to larger networks without increasing simulation complexity at this level of abstraction. 
Commercial customers are assumed not to host any behind-the-meter DER, and their reliability indices therefore equal the underlying load point values.

\textbf{Inputs and scenarios:}
The case study uses Los Angeles County as the representative location. 
Residential load profiles are drawn from the ResStock dataset \cite{ResStock}, and rooftop PV generation is computed from the NSRDB irradiance data \cite{NSRBD}. 
Since ResStock internally uses NSRDB irradiance as its solar input, the load and generation data are inherently consistent, eliminating potential mismatches between demand and resource drivers. 
The adoption distributions for PV and ES, along with their correlation structures, are defined in Section~\ref{sec:adoption} and summarized in Table~\ref{tab:adoption scenario table}. 
These distributions are treated as scenario-based abstractions rather than forecasts, enabling the framework to capture a range of plausible adoption patterns.

\textbf{Simulation results:} Tables~\ref{tab:SAIFI} and \ref{tab:SAIDI} summarize the system-wide reliability indices across the sixteen PV--ES adoption scenarios. 
For reference, the baseline RBTS values without any DER are $\text{SAIFI}=0.30$~f/yr and $\text{SAIDI}=3.47$~h/yr. 
The simulated results confirm that increased adoption of PV and ES generally improves both indices, but the magnitude of improvement varies substantially across adoption patterns.

In the best-case scenario with both PV and ES highly concentrated (\{HC, HC\}), the indices reduce to $\text{SAIFI}=0.0211$~f/yr and $\text{SAIDI}=0.107$~h/yr. 
This corresponds to a reduction of approximately 93\% in interruption frequency and 97\% in interruption duration relative to the baseline, highlighting the potential value of coordinated adoption. 
At the other extreme, the limited-adoption case (\{L, L\}) produces indices nearly identical to the baseline ($\text{SAIFI}=0.3154$, $\text{SAIDI}=3.300$). 
Interestingly, in some low-adoption cases, the simulated SAIFI is slightly \emph{higher} than the baseline. 
This is because the framework accounts not only for grid-side interruptions but also for the stochastic reliability of the PV and ES subsystems themselves. 
When PV adoption is shallow and ES capacity is small, these resources may deplete quickly or fail to sustain load over the full duration of a utility outage. 
As a result, a single load point interruption can manifest as multiple end-user interruptions due to PV cycling and insufficient ES support. 
This phenomenon has been observed in our prior work \cite{karngala2023impact}, where small DER sizes led to higher interruption counts even as average interruption duration improved. 
Intermediate cases, such as the \{MF, MF\} scenario, yield $\text{SAIFI}=0.1933$ and $\text{SAIDI}=1.070$~h/yr, demonstrating substantial but less dramatic gains.

These results illustrate that the proposed framework can capture both extreme improvements and nuanced degradation mechanisms, and that its estimates are consistent with baseline expectations in low-adoption cases.

To illustrate these findings more concretely, we examine two representative scenarios in detail. 
The first highlights the maximum potential benefit under coordinated adoption, while the second reveals the heterogeneity of outcomes under diverse adoption patterns.  

To illustrate these findings more concretely, we examine two representative scenarios in detail. 
The first highlights the maximum potential benefit under coordinated adoption, while the second reveals the heterogeneity of outcomes under diverse adoption patterns.  

\textbf{Highly concentrated adoption (\{HC, HC\}):} 
In the highly concentrated scenario, both PV and ES capacities cluster at the upper end of their respective ranges. 
This leads to a dramatic improvement in reliability: $\text{SAIFI} \approx 0.021$~f/yr and $\text{SAIDI} \approx 0.107$~h/yr, corresponding to reductions of about 93\% and 97\% relative to the baseline. 
The AIF distribution is extremely tight, indicating that nearly all customers experience similar reliability gains. 
This case demonstrates the maximum potential of coordinated PV--ES adoption and shows how the framework captures significant system-wide improvements when adoption is both high and uniform.  

\textbf{Varied adoption (\{V, V\}):} 
In contrast, the varied adoption scenario exhibits wide diversity in PV and ES capacities, spanning nearly the entire feasible range. 
The resulting AIF distribution is clearly bimodal: one cluster of customers benefits significantly, achieving reliability comparable to the \{HC, HC\} case, while another cluster remains near baseline. 
The estimated mean SAIFI ($\sim$0.21~f/yr) lies between these peaks, but does not represent either group well. 
This example highlights the central contribution of the framework: single numerical indices obscure the heterogeneity of customer outcomes, whereas the distributional perspective reveals both the opportunities and the risks of uneven DER adoption.

\begin{table}[!t]
\caption{SAIFI for Joint scenarios (Baseline = 0.3 f/yr)}
\label{tab:SAIFI}
\centering
\begin{tabular}{ccccc}
\hline
\multirow{2}{*}{PV} & \multicolumn{4}{c}{\bf ES} \\ \cline{2-5}
 & \textbf{Limited} & \textbf{Varied} & \textbf{Median} & \textbf{Highly Conc.} \\
\hline
\textbf{Limited}        & 0.3154 & 0.3150 & 0.3129 & 0.3127 \\
\hline
\textbf{Varied}         & 0.3047 & 0.2053 & 0.2167 & 0.1540 \\
\hline
\textbf{Median}         & 0.3018 & 0.2074 & 0.1933 & 0.0987 \\
\hline
\textbf{Highly Conc.}   & 0.2867 & 0.1714 & 0.1603 & 0.0211 \\
\hline
\end{tabular}
\end{table}

\begin{table}[!t]
\caption{SAIDI for Joint scenarios (Baseline = 3.47 hrs/yr)}
\label{tab:SAIDI}
\centering
\begin{tabular}{ccccc}
\hline
\multirow{2}{*}{PV} & \multicolumn{4}{c}{\bf ES} \\ \cline{2-5}
 & \textbf{Limited} & \textbf{Varied} & \textbf{Median} & \textbf{Highly Conc.} \\
\hline
\textbf{Limited}        & 3.300 & 3.298 & 3.281 & 3.279 \\
\hline
\textbf{Varied}         & 2.277 & 1.591 & 1.532 & 1.244 \\
\hline
\textbf{Median}         & 1.993 & 1.327 & 1.070 & 0.577 \\
\hline
\textbf{Highly Conc.}   & 1.813 & 1.049 & 0.787 & 0.107 \\
\hline
\end{tabular}
\end{table}

\begin{figure*}
    \centering
    \includegraphics[width = \textwidth]{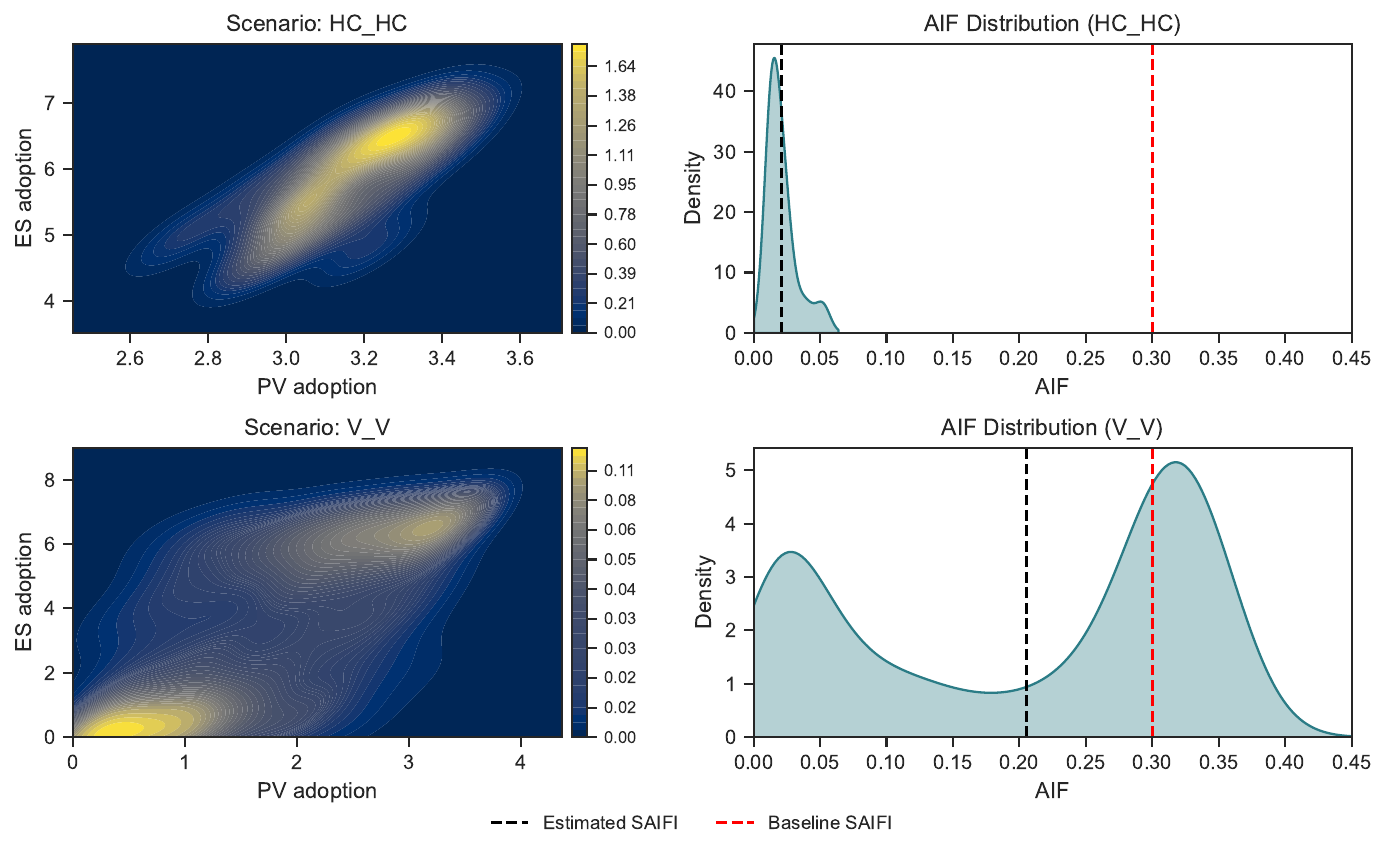}
    \caption{Joint adoption densities (left) and AIF distributions (right) for selected PV--ES adoption scenarios. 
The top row corresponds to highly concentrated adoption of PV and ES (\{HC, HC\}). 
The bottom row corresponds to varied adoption of PV and ES (\{V, V\}). 
The red dashed line marks the baseline SAIFI (0.30~f/yr) and the black dashed line marks the estimated mean SAIFI for each scenario.}
    \label{fig:custom_scenario}
\end{figure*}

\begin{table}[!t]
\caption{Representative adaptive Monte Carlo convergence statistics across PV--ES adoption scenarios. Reported cases correspond to the fastest, median, and slowest scenarios.}
\label{tab:runtime}
\centering
\begin{tabular}{lcc}
\hline
\textbf{Scenario} & \textbf{Samples} & \textbf{Runtime (s)} \\
\hline
Fastest (L, L)    & 30    & 27   \\
Median (MF, V)    & 275   & 106  \\
Slowest (HC, V)   & 2000  & 1520 \\
\hline
\end{tabular}
\end{table}

\textbf{Computational efficiency and scalability:}
The adaptive Monte Carlo framework converged in a median of approximately 275 samples, corresponding to a runtime of about 2~minutes on a standard workstation. The fastest scenarios required as few as 30 samples (27~s), while the most challenging scenario reached the prescribed maximum of 2000 samples (about 25~minutes). A summary of representative convergence statistics is provided in Table~\ref{tab:runtime}. This adaptive strategy therefore avoids the inefficiencies of fixed-sample simulations, achieving precision targets with tractable runtimes across all cases. At the first abstraction layer considered in this paper, the computational cost depends primarily on the number of samples and is independent of the network size, since all load points share uniform baseline reliability. For more detailed studies, a distribution system can be partitioned into a small number of reliability regions, each representing a set of load points with homogeneous reliability characteristics. The computational cost then scales with the number of regions rather than the total number of load points, which preserves tractability even for large feeders. Moreover, the batch-based sampling architecture ensures that the framework remains amenable to parallelization, supporting scalability for realistic distribution system applications.

% \begin{figure*}
%     \centering
%     \begin{subfigure}[b]{0.49\textwidth}
%         \centering
%         \includegraphics[width=\textwidth]{Figures/AIF_KDE.pdf}
%         \caption{SAIFI distributions for various adoption scenarios.}
%         \label{fig:AIF_results}
%     \end{subfigure}
%     \hfill
%     \begin{subfigure}[b]{0.49\textwidth}
%         \centering
%         \includegraphics[width=\textwidth]{Figures/AID_KDE.pdf}
%         \caption{SAIDI distributions for various adoption scenarios.}
%         \label{fig:AID_results}
%     \end{subfigure}
%     \caption{Each subplot is for a specific PV adoption and each distribution is for a specific ES adoption as indicated in the legend.}
%     \label{fig:saifi_saidi}
% \end{figure*}

\textbf{Uncertainty in input distributions:} 
The reliability estimates reported here are subject to both statistical and distributional uncertainty. 
Statistical uncertainty is controlled through the adaptive Monte Carlo procedure, which enforces confidence interval half-widths of 0.005 for SAIFI and 0.1~h/yr for SAIDI. 
Distributional uncertainty arises from the use of scenario-based adoption models and correlation coefficients, which are intended as plausible sensitivity cases rather than forecasts. 
These results should therefore be interpreted as sensitivity analyses around plausible adoption pathways. 
Explicit quantification of distributional uncertainty remains an important direction for future work.

\section{Conclusion}
This paper introduced a bottom-up probabilistic framework for predictive reliability assessment of distribution systems with residential DERs. By modeling adoption at the micro level and explicitly incorporating PV and ES component reliability, the framework captures distributional variability in reliability outcomes rather than relying solely on system-average indices. Application to the modified RBTS Bus 4 system demonstrated that conventional indices such as SAIFI and SAIDI mask significant heterogeneity in customer-level experiences when DERs are present.  

Numerical results showed that high DER adoption can reduce SAIFI and SAIDI by more than 90\% compared to the baseline, while shallow or uneven adoption can in some cases worsen interruption frequency. These insights highlight the importance of assessing not only mean reliability improvements but also the spread of outcomes across different customer configurations. The findings provide utilities and regulators with a robust way to anticipate the reliability impacts of DER adoption pathways and to design policies that capture their full value.  

Future work will extend this framework to larger, more detailed distribution systems by incorporating multiple reliability regions, empirical adoption datasets, and resilience-focused scenarios. These enhancements will enable more comprehensive planning studies that align with the rapid growth of residential solar and storage.  
 
% or


% \newpage
% \appendices
% \section{}\label{apen:cov}
% \input{appendix_1}

% % you can choose not to have a title for an appendix
% % if you want by leaving the argument blank
% \section{Drawing samples from multivariate distributions}
% \input{appendix_2}

% % use section* for acknowledgment
% \section*{Acknowledgment}


\ifCLASSOPTIONcaptionsoff
  \newpage
\fi

\ifCLASSOPTIONcaptionsoff
  \newpage
\fi
\bibliographystyle{IEEEtran}
\bibliography{IEEEabrv,bib}

\end{document}